\documentclass[preprint,review,12pt]{elsarticle}

\usepackage[T1]{fontenc}
\usepackage{amssymb}

\usepackage{bm}
\usepackage{color}
\usepackage{mhchem}

\newcommand{\li}{$^{7}$Li}
\newcommand{\si}{$^{29}$Si}
\newcommand{\ca}{$^{13}$C}

\newcommand{\polya}{polysilylcarbodiimide}
\newcommand{\polyb}{polysilazane}

\newcommand{\slr}{$T_1^{-1}$}
\newcommand{\ssr}{$T_2^{-1}$}
\newcommand{\slrr}{$T_{1\rho}^{-1}$}
\newcommand{\slrrs}{$T_{1\rho,s}^{-1}$}
\newcommand{\slrrl}{$T_{1\rho,l}^{-1}$}

\journal{The Journal of Power Sources}

\begin{document}

\begin{frontmatter}



\title{Li dynamics in carbon-rich polymer-derived SiCN ceramics probed by NMR}


\author[1]{Seung-Ho Baek\corref{cor1}}
\ead{sbaek.fu@gmail.com}
\author[2]{Lukas Mirko Reinold}
\author[2]{Magdalena Graczyk-Zajac}
\author[2]{Ralf Riedel}
\author[1]{Franziska Hammerath}
\author[1,3]{Bernd B\"uchner}
\author[1]{Hans-Joachim Grafe}
\address[1]{IFW-Dresden, Institute for Solid State Research,
PF 270116, 01171 Dresden, Germany}
\address[2]{Technische Universit\"at Darmstadt, Fachbereich Material- und 
Geowissenschaften, Fachgebiet Disperse Feststoffe, Jovanka-Bontschits-Stra{\ss}e 2, 64287 
Darmstadt, Germany}
\address[3]{Institut f\"ur Festk\"orperphysik, Technische Universit\"at 
Dresden, 01062 Dresden, Germany}
\cortext[cor1]{Corresponding author.  Tel:+49 (0)351 4659 801; fax: +49 (0)351 
4659 313}
\begin{abstract}
We report $^{7}$Li, $^{29}$Si, and $^{13}$C NMR studies of two different 
carbon-rich SiCN ceramics SiCN-1 and SiCN-3 derived from the preceramic polymers 
polyphenylvinylsilylcarbodiimide and polyphenylvinylsilazane, 
respectively. 
From the spectral analysis of the three nuclei at room temperature, we find 
that only the $^{13}$C spectrum is strongly influenced by Li insertion/extraction, 
suggesting that carbon phases are the major electrochemically active sites for Li storage. 
%In addition, a comparison between the \si\ spectra of SiCN-1 and SiCN-3 suggests 
%that the mixed bond tetrahedra of Si which are present in  
%SiCN-3 provide additional electrochemical storage sites.   
Temperature ($T$) and Larmor frequency ($\omega_L$) dependences of the $^7$Li 
linewidth and spin-lattice  
relaxation rates \slr\ are described by an activated law with the activation 
energy $E_A$ of 0.31 eV and the correlation time $\tau_0$ in the high 
temperature limit of 1.3 ps. 
The $3/2$ power law dependence of \slr\ on $\omega_L$  
which deviates from the standard Bloembergen, Purcell, and Pound (BPP) 
model implies that the Li motion on the $\mu$s timescale is governed by continuum diffusion 
mechanism rather than jump diffusion.  On the other hand, the 
rotating frame relaxation rate \slrr\ results suggest that the slow motion of 
Li on the ms timescale may 
be affected by complex diffusion and/or non-diffusion processes.
\end{abstract}


\end{frontmatter}


%\section{}
%\label{}



\section{Introduction}

Polymer-derived ceramics (PDCs) that consist of Si, C, O, and N possess 
novel physical, chemical, and mechanical features which can be tuned by subtle changes of 
composition and/or microstructure as well as of processing conditions,
finding applications in a variety of fields such as  
fibers, brakes for vehicles, sealants, coatings, and sensors \cite{riedel06,colombo10}.

Among the Si-based PDCs, SiCN and SiOC are promising candidates for anodes for Li-ion  
batteries to replace graphite anodes. They possess a high discharge capacity compared to that 
of graphite, thermal and chemical stability against corrosion, and cyclic 
stability during charge/discharge due to stable 3D network structure 
with amorphous 
nature \cite{wilson97,ziegler99,kolb06,feng10,mera13,liu13}. Recent studies
were focused on the carbon-rich SiCN ceramics due to 
their increased thermal stability and electrochemical performance that are 
attributed to the free carbon phase imbedded into their 
microstructure \cite{mera09,graczyk10,kaspar10,liao12,widgeon12}.   

While most of these studies concentrated on the static properties
of these materials, such as electrochemical performance, discharge
capacity, and the relation between microstructure and precursor
polymer structure, it is also very important to understand the Li
ion dynamics on various timescales, characterizing the Li mobility
and diffusion mechanism, and to get information on the storage
site of the Li.

Nuclear magnetic resonance (NMR) 
has proven to be a powerful method for probing local  
structure and Li motions in numerous Li-containing ion 
conductors \cite{grey04,heitjans05,boehmer07}.
Although there have been a couple of solid state NMR studies performed on the SiCN and  
SiOC ceramics \cite{iwamoto01,fukui10, fukui11, widgeon12}, they are mostly 
concerned with structural aspects of these  
materials using the magic-angle spinning (MAS) technique, still lacking information 
on the Li dynamics as a function of temperature and frequency. 

In this paper, we report \li, \si,  
and \ca\ NMR studies of carbon-rich SiCN PDCs,
adopting wide-line NMR method instead of its high-resolution solid state 
counterpart, providing information on the Li dynamics as well as on the 
Li storage site.  
While our data show that inserted Li ions mainly find carbons for their 
electrochemical storage sites, they also suggest that the mixed bond
tetrahedra of Si which are formed  
in \polyb-derived SiCN, act as an additional lithiation site.
The spin-lattice relaxation rates as a function of temperature and frequency 
demonstrate that the Li motion on a timescale of $\mu$s is precisely described 
by an activated law $\tau_c=\tau_0\exp(E_A/k_BT)$.  

\section{Sample preparation and experimental details}

Two SiCN ceramics derived from \polya\ (HN1) and \polyb\ (HN3) (in the following
denoted by SiCN-1 and SiCN-3, respectively)
have been synthesized, as described in detail in Refs. \cite{widgeon12,reinold13}. 
The elemental composition of the ceramics with regard to their carbon, 
nitrogen, oxygen, chlorine and hydrogen content was measured. The carbon 
amount was determined by a combustion analysis with a carbon analyzer Leco 
C-200, the nitrogen and oxygen content by hot gas extraction with a Leco 
TC-436 N/O analyzer (Leco Corporation, Michigan, USA). The chlorine and the 
hydrogen content were measured at the Mikroanalytisches Labor Pascher 
(Remagen, Germany). The silicon content was calculated as the difference of 
the above mentioned elements to 100 \%.

The particle size distribution of the ceramic powders was measured with an 
analysette 22 COMPACT (Fritsch, Germany) which is working in a measurement 
range of 0.3 to 300 $\mu$m. The measurements were performed in ethanol at constant 
stirring and ultra sonic treatment.  
The specific surface area (SSA) and porosity of the samples were determined 
from the nitrogen adsorption and desorption isotherms with an Autosorb-3B (Quantachrome 
Instruments, USA) at 77 K using the Brunauer-Emmett-Teller equation and the 
Barret-Joyner-Halenda method, respectively.

The ceramic powders were mixed with 7.5 wt\% of CarbonBlack Super 
P\textsuperscript{\textregistered}  (Timcal Ltd., Switzerland) and 7.5 
wt\% polyvinyilidenfluoride (PVDF, SOLEF, Germany)
dissolved in N-methyl-2-pyrrolidone (NMP, BASF, Germany).
The slurry was spread on a glass plate and dried
at 40 $^\circ$C for 24 h, scratched off and ground. 
Approximately 100 mg of the as prepared powder was pressed uniaxially with 30 
kN for 5 min to obtain pellets with a diameter of 10 mm and a thickness of 
roughly 0.8 mm.  
The pellets were dried under vacuum in a 
Buchi\textsuperscript{\textregistered} Glas Oven at 80 $^\circ$C for 24 h and 
transferred into a glovebox (MBraun, Germany).

All together six samples have been prepared for NMR probing. One pellet of 
each ceramic was first fully lithiated and afterwards delithiated following 
the later described procedure (in the following denoted as SiCN-1b and 
SiCN-3b). A second pellet of each ceramic was only fully lithiated (in the 
following denoted as SiCN-1a and SiCN-3a) and third pellet of each ceramic was 
prepared in the same manner as the above mentioned samples, however this 
samples was not tested electrochemically (in the following denoted as SiCN-1 
as-prepared and SiCN-3 as-prepared). 

Electrochemical testing was done in a two electrode 
Swagelok\textsuperscript{\textregistered} type cell with lithium foil (99.9\% 
purity, 0.75 mm thickness, Alfa Aesar, Germany) as counter/reference 
electrode, 1 M LiPF$_6$ in EC:DMC 1:1 wt\% (LP30, Merck, Germany) as 
electrolyte and a glass fiber separator (QMA, 
Whatmann\textsuperscript{\texttrademark}, UK). 
Additionally a polypropylene separator (Celgard 2500, Celgard, USA) was placed 
between the glass fiber separator and the pellet to avoid contamination of the 
pellets with glass fibers. 
Lithiation and
delithiation was performed with a VMP3 multipotentiostat (Biologic Science 
Instruments, France) at a current of 3.72 mA/g$^{-1}$, which is equivalent to 
a C-rate of C/100 in terms of the theoretical capacity of graphite.  
Measurements were performed at 25 $^\circ$C and 
voltage limits were set to 0.005 V and 3 V vs. Li/Li$^+$.
Afterwards the cells were disassembled in the glove box.
The pellets were washed with DMC to remove the salt
of the electrolyte and ground to a powder for NMR measurements.

The NMR measurements were performed using Tecmag Fourier Transform (FT) pulse 
spectrometer.
\li, \si, and \ca\ NMR spectra were obtained using 
a spin-echo pulse sequence ($\pi/2-\tau-\pi$) which is more advantageous than 
the free induction decay (FID)  
method since it effectively eliminates spurious signals. For the \si\ and \ca, due 
to the weak signal intensity  
associated with their low natural abundances and the long spin-lattice relaxation 
time $T_1$, their spectra were acquired only at room temperature for two 
phases (SiCN-1 as prepared and SiCN-1a) and three 
phases (SiCN-3 as prepared, SiCN-3a, and SiCN-3b). For these two   
nuclei, the NMR spectra each were averaged more than 4000 scans, with a typical 
$\pi/2$ pulse length of 4 $\mu$s and a repetition time of 40 s.

To probe the Li diffusion parameters, $^7$Li NMR spectra and relaxation rates were  
measured as a function of temperature in the range of 80--420 K. 
The spin-lattice relaxation rates  
\slr\ in the laboratory frame and \slrr\ in the rotating frame were measured 
to study the Li motions on $\mu$s and ms timescales, respectively. 
For the \slr\ measurements, the saturation recovery method was employed and $T_1$ 
was obtained by fitting the relaxation of the nuclear magnetization $M(t)$ to 
a single exponential  function, $1-M(t)/M(\infty)=a\exp(-t/T_1)$ where $a$ is a 
fitting parameter. 

In this study, since SiCN-1 and SiCN-3 exhibit very similar electrochemical 
performance as well as NMR results at 116.64 MHz, 
the detailed frequency dependences of the linewidth and relaxation rates have 
been made only on SiCN-3a.  

\section{Experimental results and discussion}

\subsection{Characterization}

The results of elemental composition are shown in Table \ref{tab:composition}. 
The table also  
includes the amount of free carbon within the ceramics calculated according to 
the equation, 
\begin{equation}
\label{eqn:ea}
\text{wt\%free C}=\frac{\left(x-1+\frac{y}{2}+\frac{3z}{4}\right)\cdot 
M_\text{C}}{M_\text{Si}+x\cdot M_\text{C}+y\cdot M_\text{O}+z\cdot M_\text{N}}\cdot100 
\end{equation}
taken from Ref. \cite{ahn11} with $x$, $y$, $z$ being taken from the empirical 
formula SiC$_x$O$_y$N$_z$ and $M_\text{C}$, $M_\text{Si}$, $M_\text{O}$ and 
$M_\text{N}$ being the molar mass of the corresponding elements. 
Both samples exhibit a high amount of free carbon and only little impurities 
of oxygen. The residual chlorine measured in the samples is due to end groups 
of chlorine at the synthesized polymers and leftovers of the byproduct 
trimethylchlorosilane of the synthesis reaction.

The evaluation of particle size distribution measurements lead to D$_{50}$ values 
of 11.6 $\mu$m for SiCN-1 and 10.6 $\mu$m for SiCN-3, respectively. 
Both SiCN ceramics demonstrate a non-porous character with a SSA lower than 
10 m$^2$g$^{-1}$ and 15 m$^2$g$^{-1}$ for SiCN-1 and SiCN-3, respectively.

\subsection{Electrochemical results}

\begin{figure}
\centering
\includegraphics[width=0.7\linewidth]{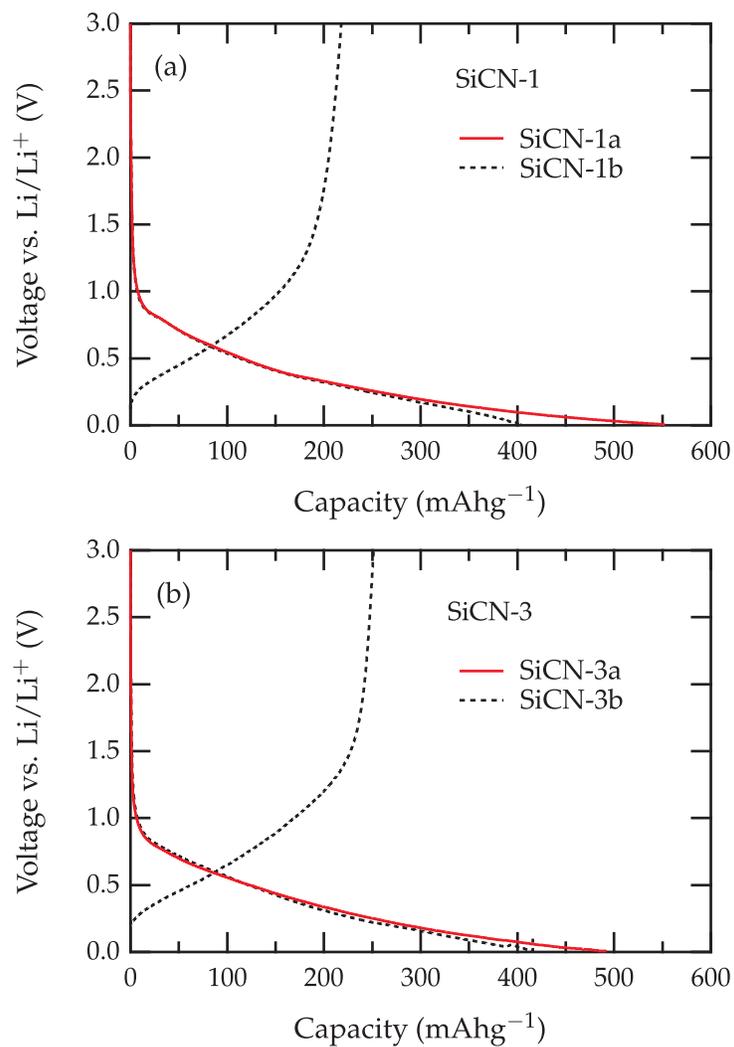}
\caption{\label{fig:capacity} Voltage vs. Li/Li$^+$ over capacity curves of 
the investigated samples. Lithiation curves of SiCN-1a and SiCN-3a are given 
by continuous red lines. The dashed black lines display lithiation and 
delithiation curves of SiCN-1b and SiCN-3b, respectively.  } 
\end{figure}

Table \ref{tab:electrochem} summarizes the lithiation and delithiation capacities of the 
investigated samples and the coulombic efficiency $\eta$ of the samples SiCN-1b and 
SiCN-3b calculated as the ratio of delithiation capacity to lithiation 
capacity times 100.  
The corresponding voltage over 
capacity plots are shown in Fig. \ref{fig:capacity}. Despite the low 
current applied for lithiation and delithiation, capacities
for both materials are below the achieved capacities of
printed electrodes. This discrepancy can be explained by
the much higher thickness of the pellets needed for a
suitable amount of powder for NMR probing.

\subsection{NMR spectra at room temperature}

\begin{figure}
\centering
\includegraphics[width=0.7\linewidth]{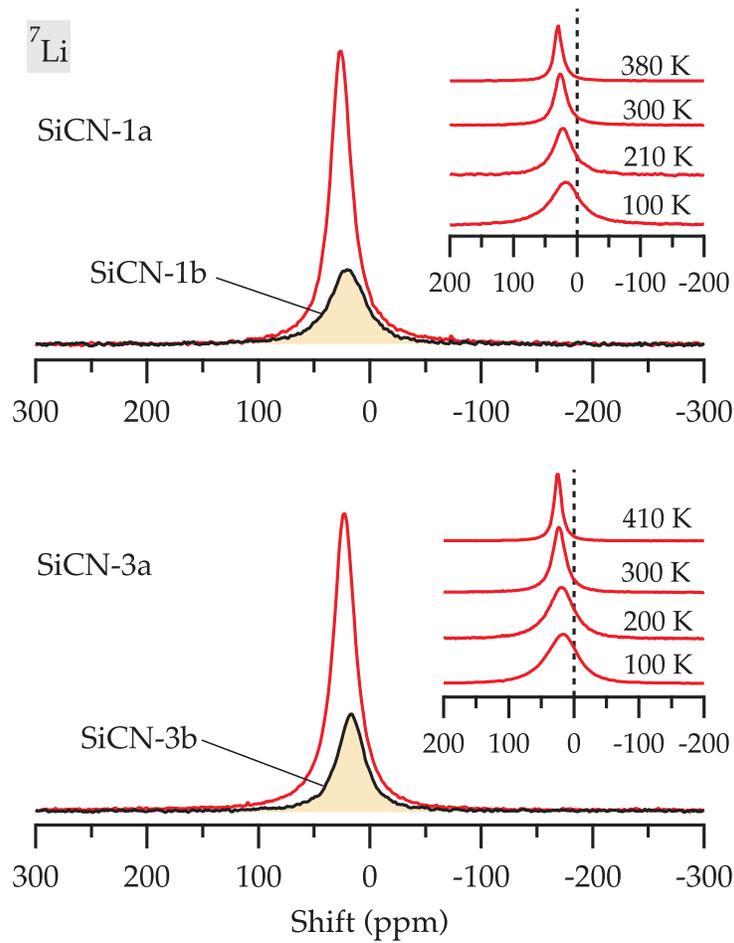}
\caption{\label{fig:sp7} $^7$Li NMR spectra of SiCN-1a and SiCN-3a samples 
taken at 300 K at 116.64 MHz.  
The $^{7}$Li spectra measured after Li extraction are compared. Clearly, the spectral 
weight at larger shift is significantly depleted after discharging. The $^7$Li 
spectra at chosen temperatures are shown in the inset.
} 
\end{figure}

\begin{figure}
\centering
\includegraphics[width=0.7\linewidth]{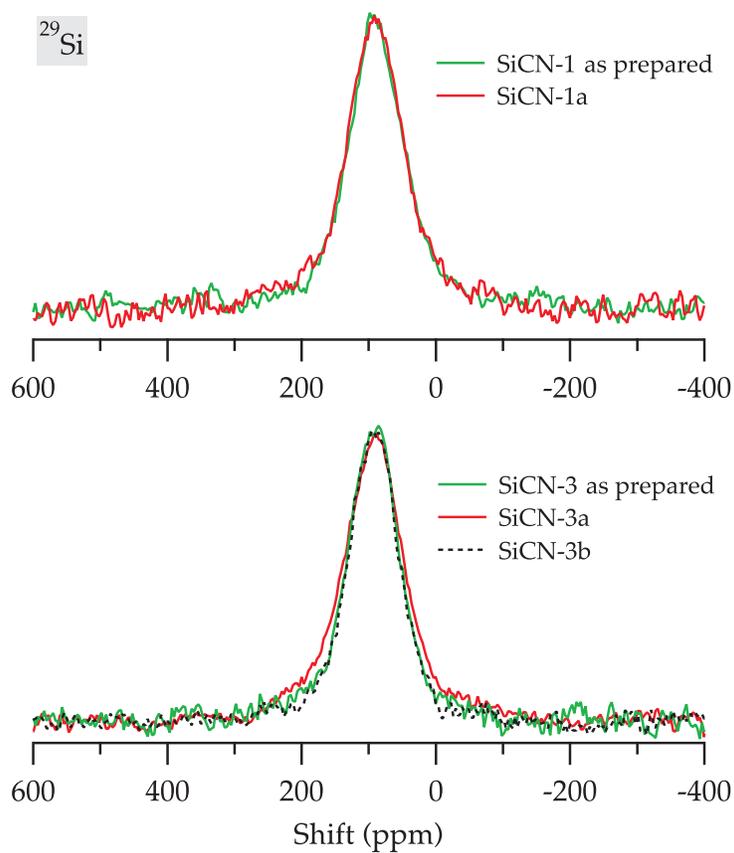}
\caption{\label{fig:sp29} $^{29}$Si NMR spectra of SiCN-1 and SiCN-3 samples 
obtained at 300 K at 59.624 MHz. While Li insertion/extraction has little influence on the 
spectrum for SiCN-1, a small broadening was identified after Li insertion for 
SiCN-3. Note that the broadening effect disappears after Li extraction.}
\end{figure}

\begin{figure}
\centering
\includegraphics[width=0.7\linewidth]{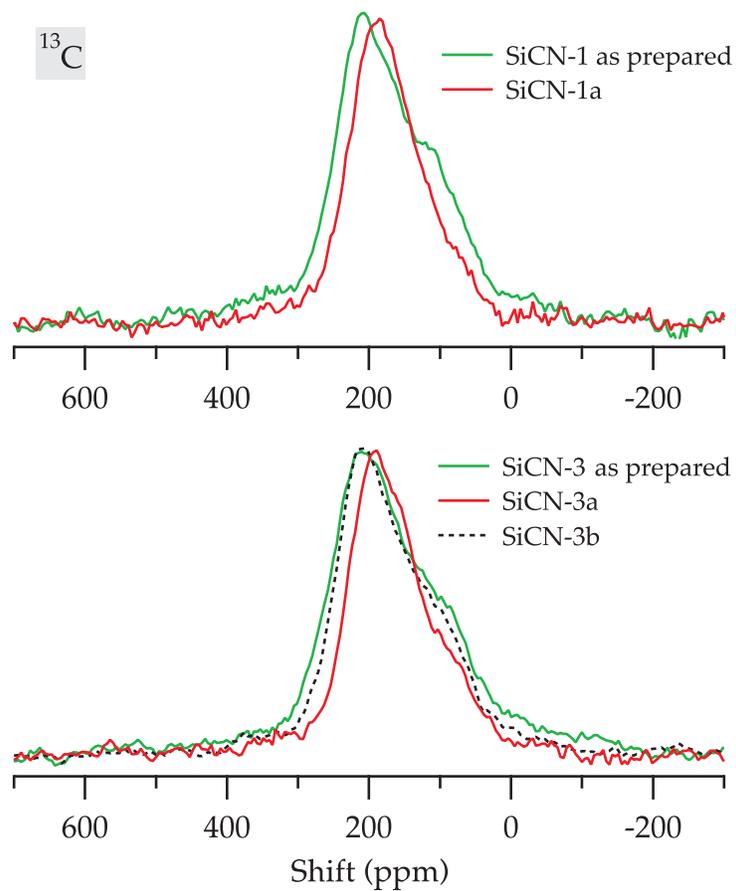}
\caption{\label{fig:sp13} $^{13}$C NMR spectra of SiCN-1 and SiCN-3 samples measured at 300 K 
and 75.476 MHz. Li insertion has a significant influence on the $^{13}$C for 
both SiCN-1 and SiCN-3 in contrast to $^{29}$Si, indicating that Li ions mostly reside 
near carbon sites regardless of the detailed local structure. }  
\end{figure}

Figures \ref{fig:sp7}, \ref{fig:sp29}, and \ref{fig:sp13} show the 
spin-echo spectra of $^7$Li,  
$^{29}$Si, and $^{13}$C, respectively, acquired at room 
temperature in an external field of 7.0494 T (i.e. Larmor frequencies of 116.64 
MHz, 59.624 MHz, and 75.476 MHz) for SiCN-1 and SiCN-3 samples.

The $^7$Li spectra shown in Fig. \ref{fig:sp7} reveal a narrow linewidth of 2 kHz 
with a small positive shift for both SiCN-1a and SiCN-3a samples. In practice, we do 
not observe any noticeable difference between the $^7$Li spectra of SiCN-1 and SiCN-3, 
suggesting that the Li dynamics is weakly sensitive to the precursor polymers.     
We also measured the $^{7}$Li spectra in the  
discharged samples. Interestingly, we find that the Li 
ions associated with smaller shift largely remain after discharging for both samples. 
This finding suggests that it is more difficult to remove Li that occupies sites 
in isotropic surroundings with less chemical shifts. 

Both samples SiCN-1 and SiCN-3 yield symmetric $^{29}$Si spectra which
are shown in Fig. \ref{fig:sp29} for the as-prepared and lithiated
samples. 
It is interesting to note that for SiCN-3 the  
linewidth increases after charging and recovers that of the as-prepared after 
discharging, while there is no difference of the spectrum after charging for SiCN-1.  
This suggests that inserted Li ions are coupled to Si, leading to a
larger distribution of the local field at the \si, i.e. a larger linewidth, in 
SiCN-3.
The fact that such a broadening does not occur in SiCN-1 
implies that local structures of Si that may act as 
electrochemical Li storage sites are formed in the SiCN ceramic derived from HN3 but 
not from HN1.

While we observed a symmetric line for $^{29}$Si, the $^{13}$C spectrum is clearly 
resolved into two in the as-prepared SiCN sample, indicating the existence of two 
inequivalent carbon sites  
or environments as shown in Fig. \ref{fig:sp13}. Since this carbon-rich SiCN 
ceramics contain a very high concentration of free carbon phases,
we attribute these two peaks to the presence of $sp^2$ and $sp^3$ 
carbons \cite{mera13,reinold13}.
In fact, a previous high-resolution $^{13}$C NMR study in SiCN revealed two 
groups of spectra associated with $sp^2$ and $sp^3$ carbons \cite{bill98}.
They are roughly separated by $\sim 110$ ppm, which indeed accounts for the 
difference between the main peak at $\sim200$ ppm and the shoulder at $\sim90$ ppm. 
Therefore, we assign  the main peak and the shoulder to $sp^2$ and $sp^3$ 
carbons, respectively. This implies that the small amount of $sp^3$ 
carbons is still present even at  
the high pyrolysis temperature (1100 $^\circ$C) in both SiCN-1 and SiCN-3. 

The insertion of Li ions affects the $^{13}$C  
spectrum significantly, reducing the anisotropy of the spectrum.
This contrasts with the $^{29}$Si spectrum which is almost intact with Li 
insertion for SiCN-1. 
The large influence of Li insertion on the $^{13}$C spectrum indicates that 
electrochemically active sites for Li storage  
are mainly carbon phases.
It would be interesting to recall a debate on the main Li storage in SiOC ceramics.   
Fukui et  al.  \cite{fukui10} suggested that the free carbon   
phase is the main site 
for Li storage, while Ahn et al. \cite{ahn11} proposed the mixed bond tetrahedra of Si 
as major Li hosting sites. 

It is known that HN1-derived SiCN-1 has no concentration of 
mixed bonds of Si unlike HN3-derived SiCN-3 \cite{ziegler99, iwamoto01}.
Thus, based on the almost identical spectra of \ca\ for both SiCN-1 and SiCN-3 samples after 
charging/discharging, one can argue that carbons are the major electrochemical  
storage site for the Li ions, regardless of the preceramic polymers. The 
fact that the whole \ca\ spectrum, which reflects two inequivalent carbon phases, 
changes after charging/discharging in a reversible way implies that all carbon 
species participate in hosting Li ions and, once binding Li, both $sp^2$ and 
$sp^3$ carbons experience similar chemical environment surrounding them.  

While the \ca\ NMR results show the main role of free carbon phases as the 
storage sites of Li, we note that the broadening of the \si\ spectrum with Li 
insertion occurs only for SiCN-3 as shown in Fig. \ref{fig:sp29}. This
suggests that the mixed bond tetrahedra of Si also act as an additional Li 
storage site, although its role is estimated to be only minor based on 
the fact that the $^7$Li spectra are almost identical for both SiCN-1 and 
SiCN-3 as shown in Fig. \ref{fig:sp7}. 

%Note that the both \si\ and \ca\ spectra of the charged SiCN-3 sample return to 
%those of the parents after Li extraction.  Taking into account the remaining 
%Li ions after discharging (see Fig. \ref{fig:sp7}),
%this result is consistent with the good reversible capacity during the 
%initial cycles in these materials.\cite{reinold13}

\subsection{Temperature dependence of $^7$Li spectra and spin-spin relaxation rate}

\begin{figure}
\centering
\includegraphics[width=\linewidth]{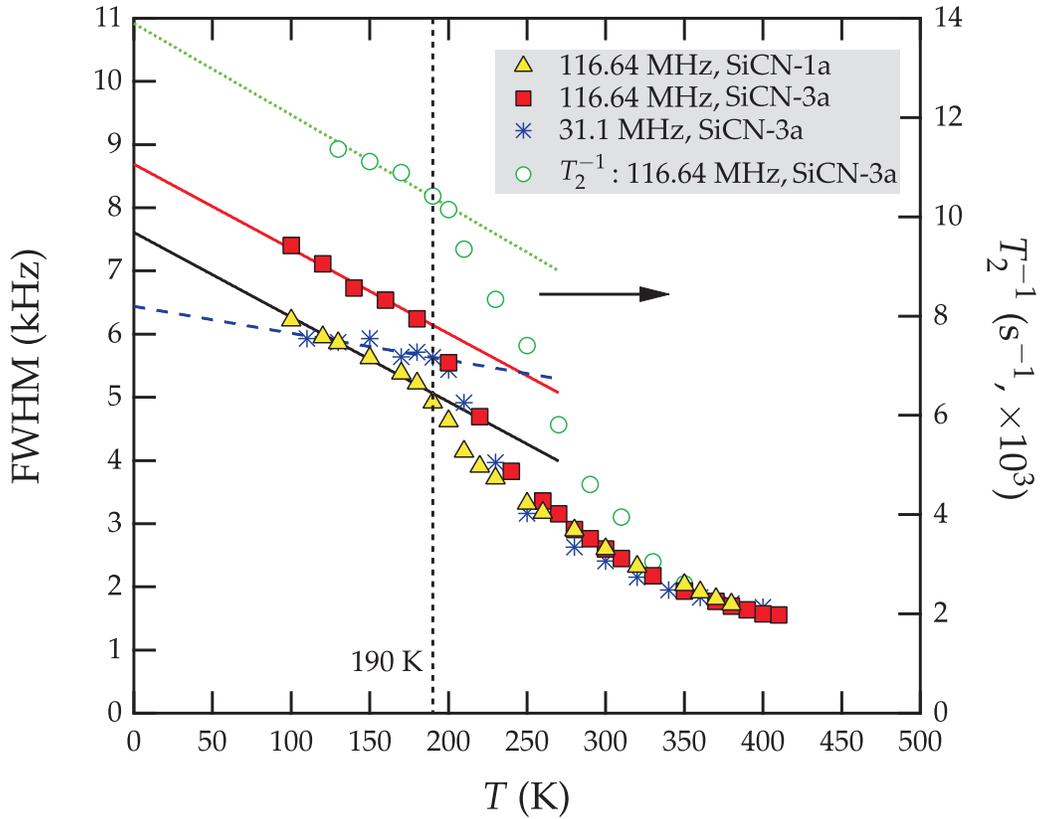}
\caption{\label{fig:fwhm}Temperature dependence of the $^7$Li NMR linewidth 
and spin-spin relaxation rate $T_2^{-1}$. The data increase rapidly with 
decreasing temperature and a  
crossover to a $T$-linear behavior below $\sim190$ 
K occurs, regardless of  
Larmor frequencies for both SiCN-1 and SiCN-3 ceramics.} 
\end{figure}

Figure \ref{fig:fwhm} shows the full width at half maximum (FWHM) $\Delta\nu$ 
and the spin-spin  
relaxation rate $T_2^{-1}$ of $^{7}$Li in lithiated SiCN samples as 
a function of temperature at two Larmor frequencies. 

The FWHM increases rapidly with decreasing temperature  
and undergoes a crossover to a  
linear $T$ behavior below $\sim190$ K. Also we find that the spin-spin 
relaxation rate $T_2^{-1}$ follows 
the same $T$-dependence as the FWHM, as expected in the slow motion regime where 
$T_2^{-1}\propto \Delta\nu$.
Although the FWHM and $T_2^{-1}$ increase 
linearly below 190 K instead of reaching a plateau which would indicate the 
rigid lattice (RL) regime,  
it should be noted that  
the linear slope below 190 K for SiCN-3 is greatly reduced at a lower Larmor frequency of 
33.1 MHz, while the data at $T>190$ K are almost independent of frequency.
Furthermore, we find that the slope for SiCN-1 and SiCN-3 at the 
same frequency of 116.64 MHz remains the same, as indicated by solid lines, 
despite their different $T$-dependence above 190 K.  
These results strongly suggest that the $T$-linear increase below 190 K could 
be an extrinsic effect being proportional to the external field strength.  
Therefore, we take 190 K as  
the onset temperature $T_\text{onset}$ of the RL regime, above which  
the hopping diffusion of Li ions takes place leading to motional narrowing of 
the $^7$Li linewidth.

In the high-temperature limit (i.e., motional narrowing limit), where the spin-spin  
relaxation rate is much smaller than the rigid-lattice linewidth i.e., 
$T_2^{-1}\ll \Delta\nu_\text{RL}$, the NMR linewidth is independent of temperature 
and $T_2^{-1}$ becomes  
proportional to the correlation time $\tau_c=\tau_0\exp(E_A/k_BT)$. Thus, the 
temperature dependence of \ssr\ usually
allows one to determine the activation energy $E_A$ and $\tau_0$ which are associated 
with the diffusive motion of Li ions. However, the motional narrowing limit is not 
accessible up to 420  
K which is the highest temperature set by our equipment, as shown in Fig. \ref{fig:fwhm}.
In this case, there is an alternative way to estimate $E_A$ using the well-known
empirical relation \cite{heitjans05},

\begin{equation}
\label{}
E_A \text{ (eV)} \sim 1.617 \times 10^{-3} T_\text{onset} \text{ (K)}.
\end{equation}

Using $T_\text{onset}=190$ K, one can obtain $E_A\sim0.31$ eV (or $\sim3600$ K) which 
indeed appears to be a reasonable value, being comparable to 0.32 eV for 
the fast 3D Li motion observed in another potential anode material \cite{kuhn11} 
Li$_{12}$Si$_{7}$, and we will  show below that the  
obtained $E_A$ is consistent with temperature and frequency dependences of the 
spin-lattice relaxation rate measurements.

\subsection{Lithium dynamics from the spin-lattice relaxation rates}

\begin{figure}
\centering
\includegraphics[width=\linewidth]{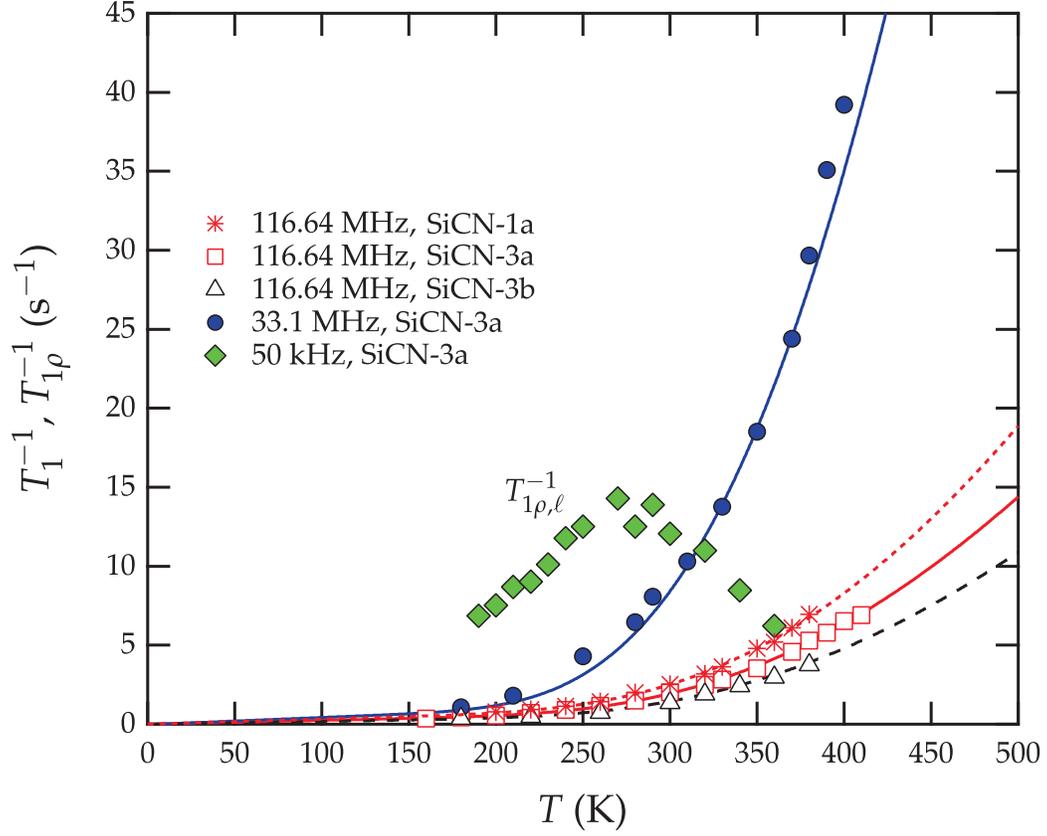}
\caption{\label{fig:invT1vsT} Temperature dependence of the $^7$Li spin-lattice 
relaxation rates \slr\ in the  
laboratory frame and \slrr\ in the rotating frame. \slrr\ (green diamond 
symbol) was measured at 50 kHz on the SiCN-3a sample.  The 
theoretical fits to the data are from Eq. \ref{eq:t1} with $E_A=3600$ K and 
$\tau_0=1.3$ ps, demonstrating the $\omega_L^{-3/2}$ dependence of \slr. }
\end{figure}

Figure \ref{fig:invT1vsT} shows the $^{7}$Li spin-lattice relaxation rate \slr\ 
as a function of temperature.   
\slr\ continues to increase with increasing temperature up to 420 K, without 
forming a maximum which is expected when the condition $\omega_L \tau_c=1$ is met from the 
Bloembergen, Purcell, and Pound (BPP) model \cite{bloembergen48},
\begin{equation}
\label{eq:bpp}
     T_1^{-1} \propto \frac{\tau_c}{1+\omega_L^2\tau_c^2}.
\end{equation}
This indicates that the system lies in the low temperature limit, i.e. 
$\omega_L \tau_c \gg 1$, over the whole temperature range investigated.  
  
Furthermore, we find that the Larmor frequency dependence of \slr\ does not 
follow $T_1^{-1}\propto \omega_L^{-2}$ which is expected from the standard BPP 
model in Eq. (\ref{eq:bpp}), 
but obeys the relation $T_1^{-1}\propto \omega_L^{-3/2}$ instead.  
In general, such a non-BPP behavior could be 
attributed to both structural disorder and Coulomb interaction \cite{meyer93}. 
In particular, the $3/2$ 
power law dependence of \slr\ on $\omega_L$ is  
often observed in disordered materials and indicates that the Li ionic diffusion occurs 
continuously (continuum diffusion) due to varying atomic distances, rather 
than jump diffusion from one site to another \cite{bjorkstam80, heitjans05}. In 
this case, \slr\ is expected to be proportional to  
$\tau_c^{-1/2}\omega_L^{n-2}$ with $n=1/2$.

\begin{figure}
\centering
\includegraphics[width=\linewidth]{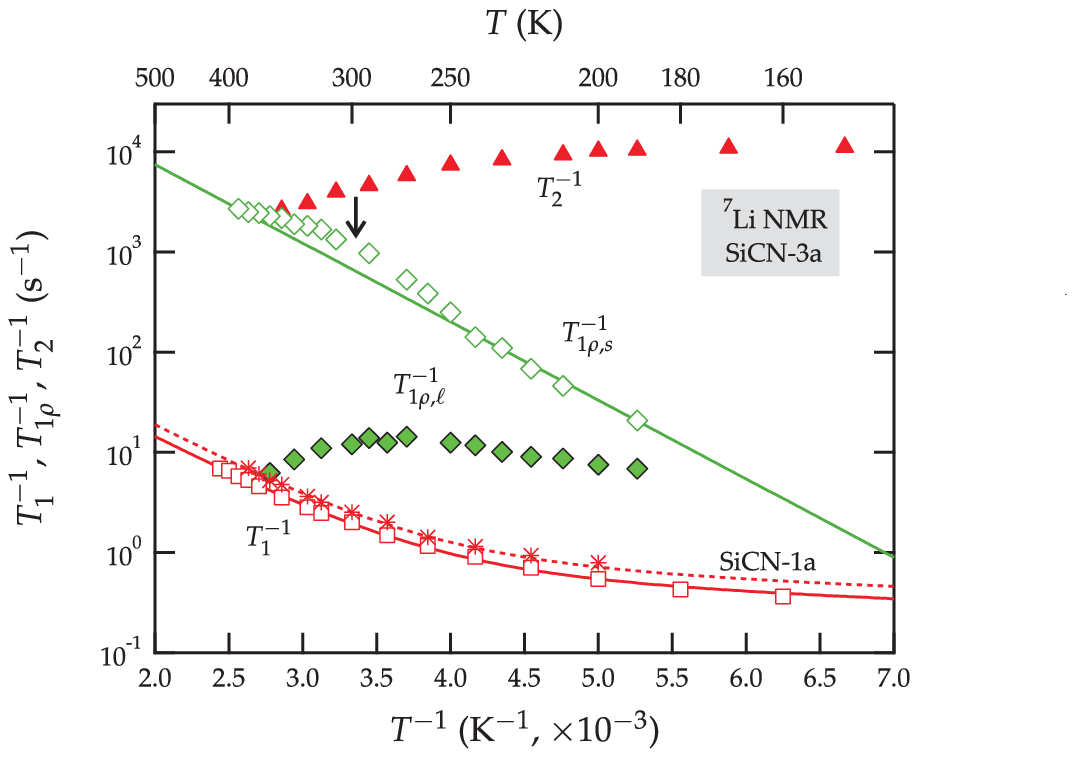}
\caption{\label{fig:invT1vsinvT} Comparison of $^7$Li relaxation rates \slr\ 
and \ssr\ at 116.64  
MHz and \slrr\ at 50 kHz as a function of inverse temperature measured  
in SiCN-3a. \slr\ of SiCN-1a at 116.64 MHz  
(red asterisks with dashed line) is almost identical to that of SiCN-3a. For 
both samples, a $T$-linear background at low temperatures superimposed on the  
diffusion-related activation behavior is clearly visible in \slr.}  

\end{figure}

In addition to the $\omega_L$-dependence, 
the data reveal the presence of a
background \slr\ which arises from contributions that are unrelated to the 
diffusion process, which is clearly revealed in the Arrhenius plot at low 
temperatures, as shown in Fig. \ref{fig:invT1vsinvT}.   
Assuming a $T$ linear background term, one can write:
\begin{equation}
\label{eq:t1}
T_1^{-1} = a_0 \frac{1}{\sqrt{\tau_c}} \omega_L^{-3/2} + c_0\omega_L^{-1/2} T,
\end{equation}
where $\tau_c\equiv \tau_0\exp(E_A/k_BT)$, and $a_0$ and $c_0$ are  
constants which are allowed to vary depending on the samples.  

We find that this formula excellently describes the temperature and  
$\omega_L$-dependence of the \slr\ data for SiCN-3a, using $E_A=0.31$ 
eV obtained above and $\tau_0=1.3$ ps. 
We constrained the 
parameters $a_0$ and $c_0$ to keep their values during the fit of the \slr\ 
data at 33.1 and 116.64 MHz for SiCN-3a, 
in order to ensure the correct $\omega_L$-dependence of \slr.  
The fitted results are shown as solid lines in Figs.  
\ref{fig:invT1vsT} and \ref{fig:invT1vsinvT}.  
For SiCN-1a and SiCN-3b, $a_0$ and $c_0$ are  
allowed to be slightly adjusted, but the $E_A$ and $\tau_0$ values obtained above still 
hold. In fact, the \slr\ of SiCN-1a is essentially the same as that of 
SiCN-3a with a small offset, as shown in Fig. \ref{fig:invT1vsinvT}.
Note that the $E_A$ corresponds to the 
activation energy in the high $T$ regime, $E_A^\text{HT}$, so that the effective 
activation energy $E_A/2$ from Eq. (\ref{eq:t1}) is equivalent to $E_A^\text{LT}$ 
in the low $T$ regime 
that satisfies the relation $E_A^\text{LT}=(1-n)E_A^\text{HT}$ with $n=1/2$ \cite{ngai92}.

In order to get further information on the slow motion of Li ions, we measured the 
rotating frame relaxation  
rate, \slrr, which is sensitive to slow motion in the kHz frequency scale. 
Since the \slr\ measurements yielded the same diffusion parameters for 
both SiCN-1a and SiCN-3a samples, we carried out the \slrr\ measurements on the 
SiCN-3a sample only. 
We find that the relaxation of the nuclear magnetization $M_\rho(t)$ in the 
rotating frame is 
strongly stretched in its initial decay, in contrast to the \slr\ 
experiments where the  
relaxation of $M(t)$ is always single exponential. This non-exponential decay of 
$M_\rho(t)$ along the rotating field is attributed to limited spin  
diffusion in which the direct relaxation rate is faster than the diffusion 
rate \cite{tse68,balzer89}. In this case, one may expect that $M_\rho(t)$ decays 
exponentially at long times 
where the diffusion rate becomes important. Indeed, we observed a 
single exponential decay at long times.   
Here, we extracted \slrr\ in two 
ways, one from taking $t$ when $M_\rho(t)=M_\rho(0)/e$ and the other from the 
fit of the exponential decay at long times, \slrrs\ and 
\slrrl, respectively.  \slrrl\ as a function of $T$ is shown in Fig. 
\ref{fig:invT1vsT}, exhibiting a clear characteristic temperature 
$T_\text{max}$ at which the spin-lattice relaxation  
rates pass through a maximum. Remarkably, $T_\text{max}$ 
is very close to 260 K at which
$2\omega_1\tau_c=1$, where $\omega_1$ is the Larmor frequency in the RF field, with the 
same diffusion parameters $E_A$ and $\tau_0$ deduced above.   
This is another evidence that the correlation time $\tau_c$ describing the Li 
ionic motion in SiCN is uniquely given by well-defined parameters. 

In contrast to the long rate \slrrl, the short rate \slrrs\ does not reveal 
a maximum as shown in Fig. \ref{fig:invT1vsinvT} similar to the case of \slr.    
This suggests that \slrrs\ is largely caused by the mechanism related to the 
fast Li motion. Indeed, its behavior in terms of inverse temperature, 
particularly at low $T$, seems to 
be described by Eq. (\ref{eq:t1}) with the same activation energy $E_A$ used 
for the \slr\ fit (green solid line in Fig. \ref{fig:invT1vsinvT}). 
Then, one may interpret that, on top of the usual activation behavior (denoted 
by the green solid line), the additional 
enhancement of \slrrs\ with respect to the green line (see the down arrow 
in Fig. \ref{fig:invT1vsinvT}) at high  
temperatures is superimposed.  This enhancement  
centered near 300 K  may be 
related to the slow diffusive motion of Li.  

While the \slrrl\ data reveal $T_\text{max}$ being
consistent with the correlation time, the values of \slrrl\ are too  
small compared to those of \slr, as shown in Fig. \ref{fig:invT1vsT}. 
This unusual feature found in   
the \slrrl\ data as well as the seeming coexistence of the two different mechanisms in 
the \slrrs\ data may suggest that the slow motion of Li ions is
given by a rather complex process compared to their fast motion which is well 
understood by the simple activation law. 

Despite the complex rotating frame relaxation rates detected on a slow 
timescale, our NMR relaxation measurements show that the fast dynamics of Li on a 
timescale of $\mu$s, which is of practical  
importance for the performance of anode materials, is   
essentially described by a ``single'' activation energy and correlation time in a wide 
temperature range. 
This is a quite remarkable finding, taking  
into consideration that the structural disorder
in amorphous materials like SiCN usually causes a distribution of either 
activation energy or correlation times or both.

\section{Conclusions}

We carried out systematic NMR studies focusing on the $^{7}$Li spectra and 
relaxation rates as a function of temperature in the two SiCN 
ceramics derived from \polya\ (HN1) and 
\polyb\ (HN3).
Our study successfully characterized the \textit{local} Li diffusion process in these 
materials by NMR linewidth and relaxation measurements. 
The dependence on Li insertion/extraction of the \si\ and \ca\ NMR spectra 
revealed that both $sp^2$ and $sp^3$ free carbons act as major 
electrochemically active sites for Li storage.  
The comparison between the \si\ spectra of SiCN-1 and SiCN-3
show that an additional path for Li storage can be formed in SiCN-3 
through the mixed bond tetrahedra of Si.

We found that the $^7$Li line narrowing occurs at 190 K, which allows an 
estimation of the activation energy of 0.31 eV.    
Apart from the slow motion regime of Li which seems to be affected by complicated 
diffusion and/or non-diffusion processes, 
the temperature and Larmor frequency dependences of the $^7$Li relaxation 
rates demonstrated that the  
Li dynamics in the polymer-derived SiCN ceramics in a wide operational temperature range 
from 200 to 400 K is governed by a very stable  
thermal activation law given by well-defined diffusion parameters.  
The stable diffusion mechanism of SiCN ceramics   
regarding temperature variation suggests that one may fine-tune 
the Li diffusion parameters for an anode material optimized for specific battery 
performance, without the need for the detailed understanding of the  
structural complexity in these amorphous materials.

\section*{Acknowledgement}
We thank Yan Gao for her help in polymer synthesis.
This research was financially supported by the DFG within the priority
program SPP1473 (Grants No. GR3330/3-1 and GR 4440/1-2) and the
collaborative research center SFB 595/A4.


\bibliographystyle{elsarticle-num} 
%\bibliography{mybib}

\newpage

\begin{table}
\centering
\caption{\label{tab:composition}
Elemental composition of SiCN-1 and SiCN-3 in wt\%. The calculation of the 
free carbon content [wt\%] was done by neglecting the amount of chlorine and hydrogen 
in the samples.} 
\begin{tabular}{lcccccccc}
\hline
Sample & Si & C & N & O & H & Cl & Free C & Empirical formula\\
%       & [wt.\%]& [wt.\%]& [wt.\%]& [wt.\%]& [wt.\%]& [wt.\%] & [wt.\%] & Formula \\

\hline
SiCN-1 & 28.70 & 54.94 & 14.73 & 0.96 & 0.60 & 0.05 & 52.7 & 
$\mathrm{Si_1C_{4.47}N_{1.03}O_{0.06}H_{0.58}}$ \\ 
SiCN-3 & 34.41 & 47.71 & 13.28 & 0.87 & 0.27 & 3.46 &43.3 & 
$\mathrm{Si_1C_{3.24}N_{0.77}O_{0.04}H_{0.22}Cl_{0.08}}$   \\ 
\hline

\end{tabular}
\end{table}

\begin{table}
\centering
\caption{\label{tab:electrochem} 
Lithiation/delithiation capacities (C$_\text{li}$/C$_\text{deli}$)
and coulombic efficiency ($\eta$) for investigated samples.}
\begin{tabular}{lccc}
\hline
Sample & C$_\text{li}$ [mAh$\cdot$g$^{-1}]$ & C$_\text{deli}$
[mAh$\cdot$g$^{-1}$] & $\eta$ [\%] \\
\hline
SiCN-1a  & 552 & - & - \\
SiCN-1b & 404 & 218 & 54 \\
SiCN-3a & 492 & - & - \\
SiCN-3b & 416 & 251 & 60 \\
\hline

\end{tabular}
\end{table}

\end{document}